\documentclass[aps,prb,superscriptaddress,twocolumn,showpacs,preprintnumbers,amsmath,amssymb]{revtex4}

\usepackage{graphicx}
\usepackage{dcolumn}
\usepackage{bm}

\begin{document}

\title{Diffractive paths for weak localization in quantum billiards}

\author{Iva B\v rezinov\'a}
\email{iva.brezinova@tuwien.ac.at}
\affiliation{Institute for Theoretical Physics, Vienna University of Technology,
1040 Vienna, Austria, EU} 
\author{Christoph Stampfer}
\affiliation{Solid State Physics Laboratory, ETH Zurich, 8093 Zurich, Switzerland}
\author{Ludger Wirtz}
\affiliation{Institute for Electronics, Microelectronics, and Nanotechnology, CNRS, 
59652 Villeneuve d'Ascq, France, EU}
\author{Stefan Rotter}
\affiliation{Department of Applied Physics, Yale University, New Haven, Connecticut, 06520, USA}
\author{Joachim Burgd\"orfer}
\affiliation{Institute for Theoretical Physics, Vienna University of Technology,
1040 Vienna, Austria, EU}  

\date{\today}

\begin{abstract}
We study the weak localization effect in quantum transport through a clean ballistic cavity with regular classical dynamics. We address the question which paths account for the suppression of conductance through a system where disorder and chaos are absent. By exploiting both quantum and semiclassical methods, we unambiguously identify paths that are diffractively backscattered into the cavity (when approaching the lead mouths from the cavity interior) to play a key role. Diffractive scattering couples transmitted and reflected paths and is thus essential to reproduce the weak-localization peak in reflection and the corresponding anti-peak in transmission. A comparison of semiclassical calculations featuring these diffractive paths yields good agreement with full quantum calculations and experimental data. Our theory provides system-specific predictions for the quantum regime of few open lead modes and can be expected to be relevant also for mixed as well as chaotic systems. 
\end{abstract}


\maketitle

\section{Introduction}
\begin{figure}[!t]
	\centering
		\includegraphics[width=0.43\textwidth]{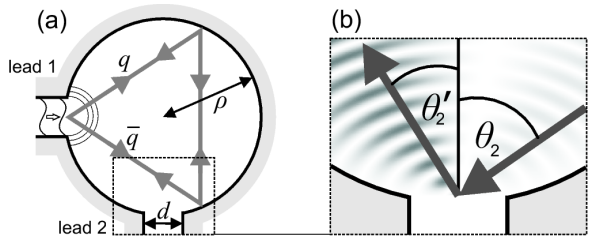}
	\caption{(a) Geometry of circular quantum billiard with radius $\rho=\sqrt{1+4/\pi}$ (in scaled units) and width of the leads $d=0.25~(\approx0.16\rho)$ with one pair of time-reversal symmetric paths $q$ and $\bar{q}$ depicted. (b) Internal diffractive backscattering at the lead, described by diffraction amplitude $v(\theta'_2,\theta_2,k)$ (plotted is $\textrm{Re}[v(\theta'_2,\theta_2=\frac{\pi}{4},k)]\cos(kr)$ with $r$ as the radial distance from the center of the lead and $k=\frac{2.5\pi}{d}$).}
\label{fig1}
\end{figure}

Weak localization (WL), the enhancement of reflection by coherent superposition of symmetry-related wave components, is an ubiquitous phenomenon of wave transport through disordered media. Experimental observations range from optics ('albedo'), to ultrasound transmission and seismic waves.\cite{ErbLenMar93,deRTouDerRouFin05,CheSuLuHua06,LarMar04} Very recently, transport of ultracold atoms through disordered optical lattices has been proposed as a new candidate for weak localization.\cite{KuhMinDelSigMul05} In mesoscopic physics WL is one of the hallmarks for phase-coherent electron transport through disordered materials.\cite{Ber84,GorLarKhm79} Its experimental observation is of particular conceptual interest in systems where disorder scattering is strongly reduced.\cite{ChaBarPfeWes94,BerKatMarWesGos94} In these \emph{ballistic} systems the WL peak in the transport resistance has been shown to be sensitive to the underlying classical regular or chaotic dynamics \cite{ChaBarPfeWes94,BarJalSto93,BarMel94} and to the presence of scattering resonances.\cite{AkiFerBirVas99}\\ 
On the most fundamental level, WL probes the particle-wave duality in transport. Identification of the relevant 'paths' in the particle-like dynamics that lead to both constructive interference in reflection and to destructive interference in transmission continues to pose a challenge. In disordered media such a mechanism can be described with the help of diagrammatic techniques that allow to sum over diffractive paths due to scattering at impurities or variations of the potential landscape.\cite{GorLarKhm79} In the ballistic regime of clean quantum dots at low temperatures where both elastic $l_e$ and inelastic mean free paths $l_i$ become large compared to the linear dot dimension, $l_{e,i} \gg \sqrt{A}$ ($A$: area of the dot), the dynamics is coherent and governed by the scattering of the electron at the (hard) boundaries of the quantum dot. Thus new concepts are required. One of these includes quantum corrections to the transport problem in analogy to the 'diffuson' and 'cooperon' propagators known from the diagrammatic techniques, assuming a static long range potential within the cavity.\cite{AleLar96} Another concept relates the conductance to a sum over \emph{classical paths} with quantum mechanical phases.\cite{BarJalSto93} In this trajectory-based approach partially time-reversed paths of almost equal length but a different number of self-intersections were invoked.\cite{RicSie02} The are considered to be responsible for WL, and were analyzed based on diffractive \cite{TakNak97} and chaotic \cite{RicSie02} scattering. The latter approach allows to reproduce the WL predictions of random matrix theory (RMT) and to fulfill the current conservation requirement.\cite{RicSie02,BroRah06,HeuMulBraHaa06,JacWhi06}\\
Identification of the relevant paths for WL in a \emph{system-specific} geometry remains an open problem. Previous theories involve an ensemble average over the sample geometry, thereby precluding the identification of system-specific paths, and an $\hbar \rightarrow 0$ limit, details of which differ from each other.\cite{BroRah06,HeuMulBraHaa06}
This complicates a direct comparison with either experiment or full quantum simulation. Moreover, completely chaotic systems are assumed, thus preventing applications to quantum dots without disorder \emph{and} chaotic scattering. For regular billiards classical paths alone are insufficient to reproduce quantum transport.\cite{BloZoz01} To uncover which mechanism for WL is at work for this class of systems, we focus in this letter on a prototype structure for a ballistic regular cavity: the circular quantum billiard with perpendicular leads (Fig.~\ref{fig1}a). The circular billiard was shown both experimentally and numerically to feature a distinctive WL effect.\cite{ChaBarPfeWes94,BerKatMarWesGos94,YanIshBur95} It does not contain any of the classical path pairs \cite{RicSie02} invoked to explain WL in transmission through ballistic chaotic cavities. We choose the geometry of the billiard and the number of open lead modes $N$ in accord with experimental values.\cite{ChaBarPfeWes94} 
%
%
By comparing exact quantum mechanical calculations \cite{RotTanWirTroBur00} with semiclassical approximations that include diffractive paths \cite{StaRotBurWir05} we are able to identify the origin of the WL dip in the transmission in terms of paths which are diffractively backscattered into the cavity and thus are intrinsically cross correlated with paths contributing to reflection.
%
\section{Semiclassical methods and quantum mechanical path spectra}
\begin{figure}[!t]
	\centering
		\includegraphics[width=0.42\textwidth]{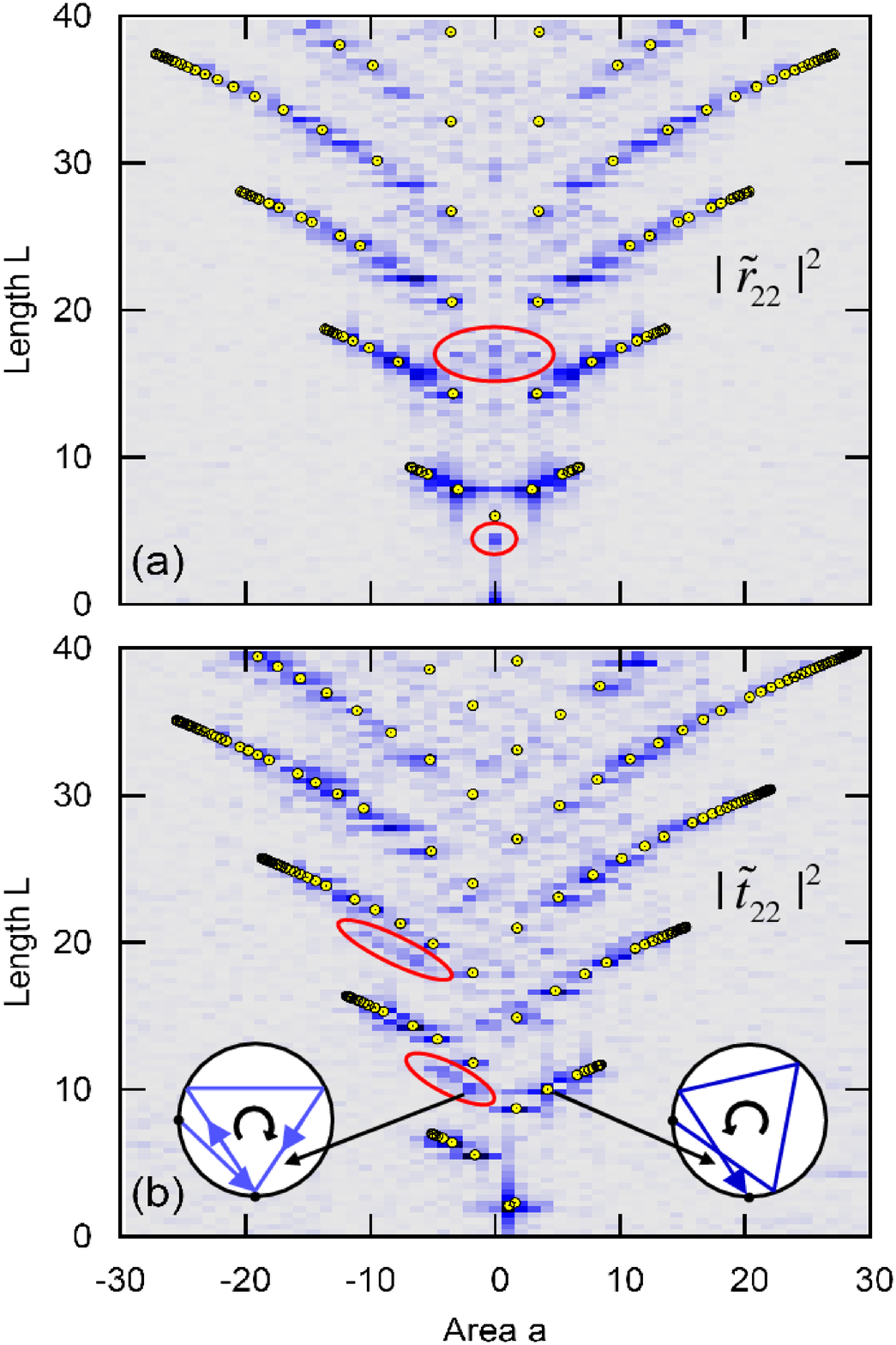}
	\caption{(Color online) 2D path length-area spectrum of the exact quantum amplitude $|\tilde{S}_{22}(L,a)|^2$ (Eq.~\ref{fure}) with integration ranges $k\in[2.2\pi/d,3.45\pi/d]$ and $B/c\in[-3,3]$ (a) for reflection 
$|\tilde{r}_{22}(L,a)|^2$ and (b) for transmission 
$|\tilde{t}_{22}(L,a)|^2$. Strength given by intensity of color, the dots mark classical paths. The pairing of classical and peudo-paths contributing to WL is visualized [insets in (b)] and examples of diffractive contributions by pseudo-paths are highlighted (red markings).}
\label{fig2}
\end{figure}

Ballistic transport is described by a wave number ($k$) and magnetic field ($B$) dependent Hamiltonian $S$ matrix with elements $S^{(j,i)}_{n,m}(k,B)$ for scattering from mode $m$ in lead $i$ to mode $n$ in lead $j$. For a two-terminal device, the transmission amplitudes from lead 1 to lead 2 are denoted as $t_{nm}=S^{(2,1)}_{n,m}$ while the reflection amplitudes back into lead 1 are denoted as $r_{nm}=S^{(1,1)}_{n,m}$. The conductance is given by the Landauer formula 
\begin{eqnarray}
g(k,B)=\frac{2e^2}{h}T(k,B)=\frac{2e^2}{h}\sum_{n=1}^{N}\sum_{m=1}^{N}|t_{nm}(k,B)|^2,
\label{eq:landcond}
\end{eqnarray}
with $T(k,B)$ being the total transmission probability. The projection of the lead modes $\phi_m(y_i)$ ($y_i$ is the transverse lead coordinate) onto the constant energy propagator $G(\vec{r_j},\vec{r_i},k,B)$ determines the $S$ matrix elements. The transmission amplitudes $t_{nm}(k,B)$, e.g.~at $B=0$, are given by
\begin{eqnarray}
t_{nm}(k,B=0)=-i\sqrt{k_{x_2,n}k_{x_1,m}} \nonumber\\
\times \int dy_2\int dy_1 \,\phi^*_n(y_2)
\,G(\vec{r}_2,\vec{r}_1,k,B=0)|_{x_1,x_2}\,\phi_m(y_1)\,,
\label{eq:tmn}
\end{eqnarray}
where $k_{x_1,m}$ ($k_{x_2,n}$) is the longitudinal wave number and $x_1$ ($x_2$) is the longitudinal coordinate in lead $1$ (lead $2$). 
Here and in the following, we use atomic units ($\hbar=|e|=m_{eff}=1$).\\
Standard semiclassical approximations (SCAs) to Eq.~\ref{eq:tmn} proceed in two steps. First, the quantum mechanical Green's function $G(\vec{r}_j,\vec{r}_i,k,B=0)$ is replaced by its semiclassical limit,
\begin{eqnarray}
G^{\rm SCA}(\vec{r}_j,\vec{r}_i,k,B=0)&=&\frac{2\pi}{{(2\pi i)}^{3/2}}\sum_{q:\vec{r_i}\rightarrow \vec{r_j}}\sqrt{|D_q(k)|} \nonumber\\&\times&
\exp{\big[iS_q(k)-i\frac{\pi}{2}\mu_q\big]}\,,
\label{eq:gsc}
\end{eqnarray}
which contains a sum over \emph{classical} paths $q(\vec{r_i}\rightarrow \vec{r_j})$ connecting the leads $i$ and $j$. The weight of each path is given by the deflection factor $D_q(k)$ which is a measure for the divergence of nearby trajectories. The phase is given by the classical action $S_q(k)$ which, for a constant potential in the interior of the dot, reduces to $S_q(k)=kL_q$ ($L_q$ is the length of the path $q$). In our semiclasssical calculations, we introduce a small magnetic field $B$ via the Aharonov-Bohm phase, $S_q(k,B)=kL_q+Ba_q/c$ ($a_q$ the directed enclosed area of the path $q$). 
The Maslov index $\mu_q$ is determined by the topology of the classical trajectory. 
In the second step, the integrals over the transverse lead coordinates in Eq.~\ref{eq:tmn} are evaluated approximatively invoking small de Broglie wavelengths (large $k$). The most common approximation is the stationary phase approximation (SPA) which enforces momentum conservation at the lead cavity junction.\cite{BarJalSto93} Going beyond the SPA, diffraction effects have been introduced on the level of the Kirchhoff diffraction approximation,\cite{SchAlfDel96} the Fraunhofer diffraction approximation (FDA),\cite{WirTanBur97} or the 'uniform theory of diffraction' (UTD),\cite{SiePavSch97} \cite{KouPat74} which is based on the 'geometrical theory of diffraction' (GTD).\cite{Kel62} For all of these diffraction approximations the transmission amplitudes take the following form,
\begin{eqnarray}
t_{nm}^{\rm SCA}(k,B)&=&-\frac{1}{\sqrt{2\pi i}}\sqrt{k_{x_2,n}k_{x_1,m}}\,\sum_{q} c_n(\theta_2,k)\sqrt{|D_q(k)|}\nonumber \\&\times&
\exp{\Big[iS_q(k,B)-i\frac{\pi}{2}\mu_q\Big]}c_m(\theta_1,k)\,,
\label{eq:tmndsc}
\end{eqnarray}
where all injection (ejection) angles $\theta\in(-\pi/2,\pi/2)$ contribute. The classical paths $q$ with specular reflections at the boundaries get weighted by the diffraction amplitudes $c_m(\theta_1,k)$ [$c_n(\theta_2,k)$] corresponding to the angle $\theta_1$ ($\theta_2$) at which they enter (exit) the cavity.
By improving the approximations used for solving the diffraction integrals, an improvement of the accuracy of the standard SCA can be achieved.\cite{WirTanBur97} In all semiclassical results presented below we evaluate the diffraction integrals using the GTD (extended by the UTD) to obtain analytical approximations for the diffraction amplitudes $c_m(\theta,k)$ featuring much better accuracy than within the FDA. In spite of this considerable improvement, the standard SCA does not reproduce quantum transport properly (even if one solves the diffraction integrals numerically exactly, see Ref.~\onlinecite{BloZoz01}). In the following we give qualitative as well as quantitative arguments for this failure. As common to all standard SCAs, the paths $q$ contributing to the transport amplitudes in Eq.~(\ref{eq:tmndsc}) are purely classical in the interior of the cavity connecting the entrance and the exit lead. We identify the missing pseudo-paths which are not contained in Eq.~(\ref{eq:tmndsc}) but essential for weak localization.\\
WL appears as a dip in the $k$ averaged transmission probability $\langle T \rangle_{\Delta k}$  and as a peak in the averaged reflection probability $\langle R \rangle_{\Delta k}$ near $B=0$. 
The full anticorrelation, $\delta \langle R \rangle_{\Delta k}=-\delta \langle T \rangle_{\Delta k}$,
highlights the fundamental issue of semiclassical descriptions in terms of interfering paths or, more generally, particle-wave duality in WL: Pairs of time reversal symmetric paths, $q(\vec{r}_1\rightarrow\vec{r}_1)$ and 
$\bar{q}(\vec{r}_1\rightarrow\vec{r}_1)$, contribute to the enhanced reflection, $\delta \langle R \rangle_{\Delta k}$\,,
due to constructive interference at $B=0$. The latter follows from the fact that the phase difference $S_q(k,B)-S_{\bar{q}}(k,B)$ vanishes for time-reversal symmetry related pairs as $B\rightarrow 0$. Such pairs, however, are uncorrelated to those pairs of trajectories $q'(\vec{r}_1\rightarrow\vec{r}_2)$, $q''(\vec{r}_1\rightarrow\vec{r}_2)$ which may interfere in transmission.
Yet, $\delta \langle T \rangle_{\Delta k}$ and $\delta \langle R \rangle_{\Delta k}$ are required to be anticorrelated. Trajectories that contribute to transmission should therefore be coupled to those leading to reflection, or in other words, subsets of classically disjoint trajectories 'must know about each other'. The deterministic outcome of either reflection or transmission must therefore be replaced by a probabilistic superposition of both transmission and reflection due to particle-wave duality. Correspondingly the WL dip in transmission can not be reproduced by employing Eq.~(\ref{eq:tmndsc}) (see Fig.~\ref{fig3}c and discussion below). The failure of Eq.~(\ref{eq:tmndsc}) to properly account for the WL effect thus implies that additional quantum effects need to be included also for the propagation \emph{in the interior of the cavity}.\\
We have identified such non-local quantum correlations between trajectories for the circular billiard. Starting point is the Fourier transform of the \emph{exact} quantum amplitudes (for details see Ref.\onlinecite{RotTanWirTroBur00}) which include all paths of the full Feynman propagator, 
\begin{eqnarray}
\tilde{S}^{(j,i)}_{n,m}(L,a)=\int dk\int dB\, e^{-i(kL+\frac{B}{c}a)}S^{(j,i)}_{n,m}(k,B).
\label{fure}
\end{eqnarray}
The Fourier conjugate variables to $k$ and $B$ are the length $L$ and the directed area $a$ enclosed by the corresponding path.\cite{SchAlfDel96,WirTanBur97,RotTanWirTroBur00,StaRotBurWir05} The resulting length-area spectra for both transmission and reflection provide unbiased information on the (non) classical properties of the whole set of paths of length $L$ and enclosed area $a$ entering Feynman's propagator, in particular of non-local correlations between reflection and transmission without invoking any semiclassical limit. 
The two-dimensional length-area spectra for reflection (Fig.~\ref{fig2}a for $|\tilde{r}_{22}(L,a)|^2$) and transmission (Fig.~\ref{fig2}b for $|\tilde{t}_{22}(L,a)|^2$) of the circular billiard, calculated with the help of the modular recursive Green's function method\cite{RotTanWirTroBur00} (MRGM), can directly be compared to paths entering the semiclassical approximation. Paths within the standard SCA are classical between entering and exiting the circular cavity and are marked by dots in Fig.~\ref{fig2} (a) and (b).\\
Maximal constructive or destructive interference at $B=0$ requires pairs of paths with (almost) identical length (on horizontal lines of fixed $L$) but symmetrically placed on both sides of the $a$ axis.
The classical length-area spectrum for \emph{reflection} (marked by dots in Fig.~\ref{fig2}a) features such symmetric branches with respect to the $a=0$ axis and accounts, to a considerable amount, for the constructive backscattering.
By contrast, classical paths contributing to transmission (marked by dots in Fig.~\ref{fig2}b)) lie on branches on opposite sides of the $a$ axis vertically displaced relative to each other and thus cannot produce a WL dip. New diffractive paths are required to fill the gap and provide approximately symmetric branches. Two families of the shortest of these diffractive contributions near $L\approx10$ and $20$ are highlighted. \\
It is now instructive to identify the origin of the paths restoring the WL dip in transmission by analyzing the pathlength and area distribution. For example, members of the shortest non-classical branch are composed of the shortest direct classical transmission path $q_1(\vec{r}_1\rightarrow\vec{r}_2)$ that diffractively backscatters at the exit lead with amplitude $v(\theta'_2,\theta_2,k)$ (Fig.~\ref{fig1}b) followed by a clockwise propagating polygon-shaped path $q_2(\vec{r}_2\rightarrow\vec{r}_2)$ that emerges from the exit lead and returns to it [left inset in Fig.~\ref{fig2}b]. Such paths are referred to as pseudo-paths.\cite{StaRotBurWir05} The pseudo-path $q(\vec{r}_1\rightarrow \vec{r}_2)$ in Fig.~\ref{fig2}b (left inset) destructively interferes with a classical path $q'(\vec{r}_1\rightarrow\vec{r}_2)$ that first misses the exit lead and subsequently encircles the billiard in counterclockwise orientation [right inset in Fig.~\ref{fig2}b]. (The relative phase between the classical and the pseudo-path at $B=0$ involves the difference in length, the difference in the Maslov indices and the phases from diffractive scattering, and is approximately $\pi$.) A segment of this second shortest member of the family of transmitted pseudo-paths has the shape of a triangle [left inset in Fig.~\ref{fig2}b]. Longer paths of the same family contain $n$-polygons which converge in the $n\rightarrow \infty$ limit to creeping trajectories along the circular boundary, and determine the endpoint of the branch (see Fig.~\ref{fig2}). 
This example directly illustrates how diffractive wave-scattering at the lead mouth causes the non-local coupling of transmitted and reflected trajectories: the reflected classical trajectory $q_2(\vec{r}_2\rightarrow\vec{r}_2)$ becomes a segment of the pseudo-path $q(\vec{r}_1\rightarrow\vec{r}_2)$ contributing to transmission. This interplay repeats itself for families of longer and more complex paths. The underlying mechanism at work here is reminiscent of 'Hikami boxes' \cite{hik81} by which piecewise 'classical' trajectories are linked to each other by diffractive scattering events.\\
The corresponding pseudo-path semiclassical approximation\cite{StaRotBurWir05} (PSCA)  allows to complement the present quantum analysis by an explicit summation of both the classical paths and the pseudo-paths. Within the PSCA diffractive changes of the direction in the interior of the cavity are incorporated into the Green's propagator in terms of a Dyson equation, $G^{\rm PSCA}\!\!=\!G^{\rm SCA}\!+\!G^{\rm SCA}\,V\,G^{\rm PSCA}$.
Here the unperturbed Green's function $G^{\rm SCA}$ coincides with Eq.~\ref{eq:gsc} including in addition paths that are geometrically reflected off the lead.\cite{SchAlfDel96} The perturbation $V$ represents the diffractive scatterings at the lead mouths with scattering vertices $v(\theta',\theta,k)$ for which we employ the GTD extended by UTD. The iterative solution to $i^{th}$ order in $V$ contains pseudo-paths consisting of $i+1$ segments of classical paths interconnected by $i$ diffractive scatterings. The corresponding length-area distribution of pseudo-paths resulting from PSCA (not shown) closely mirrors the respective quantum spectrum thus unambiguously establishing pseudo-paths as partners in the path pairing required for WL.
%
\section{Results for weak localization}
\begin{figure}[!t]
	\centering
		\includegraphics[width=0.45\textwidth]{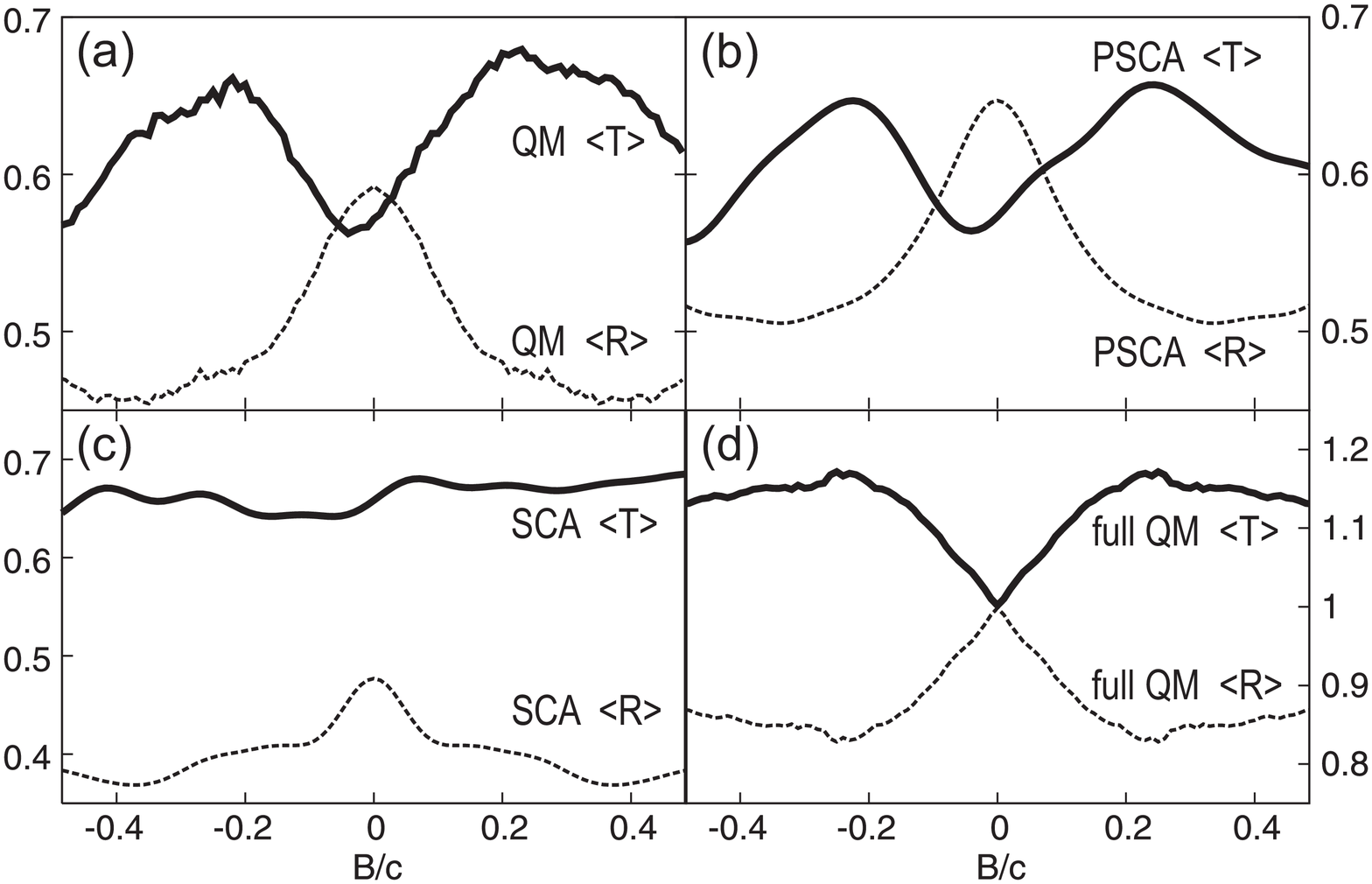}
	\caption{Weak localization (WL) in reflection and transmission for the circular quantum dot (see Fig.~\ref{fig1}) with two open modes and wave-number average $[2.2\pi/d\le k \le 2.8\pi/d]$. (a) Quantum calculation of WL with truncated set of quantum paths ($L\le L_{\rm max}=40$) (see Eq.~\ref{fure}). (b) WL calculated within the PSCA and the same truncation ($\textrm{no.\ of paths} \approx 7\times10^6$). (c) WL calculated within SCA (only classical paths) and $L\le L_{\rm max}$. Note that a dip in $T$ is completely absent. (d) Full quantum calculation (no truncation).}
\label{fig3}
\end{figure}

To quantify the role of pseudo-paths in WL we have calculated their contribution within PSCA. Since only paths up to a finite number of scattering events can be numerically summed up and all classical and pseudo-paths up to the same length must be included, the summation must be truncated for technical reasons beyond a maximum pathlength $L_{\rm max}$. This limitation, however, does not impede a \emph{quantitative} comparison with the full quantum calculation since the same truncation can be imposed on the true quantum paths (see Fig.~\ref{fig2}) by an inversion of the Fourier transform (Eq.~\ref{fure}) with a window for the pathlength $0\le L\le L_{\rm max}$ imposed. The resulting PSCA and quantum WL dips in transmission (conductance) and WL peaks in reflection are in good \emph{quantitative} agreement with each other (Fig.~\ref{fig3}a,b). We find satisfactory agreement also on the level of the fully differential k-dependence of the mode-to-mode scattering amplitudes.\cite{BreStaWirRotBur} Overall, the deviation between PSCA and quantum mechanics in the $k$-averaged mode-to-mode transmission (reflection) as a function of the magnetic field is between $0\%$ and $5\%$ ($0\%$ and $10\%$). Residual differences between the PSCA and the quantum results can be attributed to the approximations to the amplitudes $v(\theta',\theta,k)$ and $c_m(\theta,k)$ involved which can be quantified by direct comparison with the exact quantum diffraction amplitudes.\cite{SchAlfDel96}\\ 
To underline the importance of diffractive scattering in the interior, we emphasize that the standard semiclassical approximation without pseudo-paths produces no WL dip in transmission (see Fig.~\ref{fig3}c). Pseudo-paths (see, e.g., the highlighted paths in Fig.~\ref{fig2}a) also considerably improve the agreement for the WL peak in reflection (compare Fig.~\ref{fig3}b with \ref{fig3}c).
For completeness we also show the full quantum calculation for WL without truncation (Fig.~\ref{fig3}d). The influence of longer paths beyond the truncation limit (in the present example $L_{\rm max}=40$ corresponds to 13 traversals through the billiard) manifests itself in restoring the unitarity. Note, however, that our calculations with truncation reproduce the anti-correlation $\delta \langle R \rangle_{\Delta k}\simeq-\delta \langle T \rangle_{\Delta k}$, demonstrating that anti-correlation between transmission and reflection is an intrinsic property of non-classical path correlations and not just a corollary of unitarity. We also emphasize that the truncation employed renders the shape of the WL peak effectively Lorentzian customarily associated with chaotic dynamics. In the limit $L_{\rm max}\rightarrow \infty$ the triangular shaped peak associated with regular dynamics \cite{BarJalSto93,AkiFerBirVas99} is recovered. This result closely mirrors the experiment,\cite{ChaBarPfeWes94} where a transition from a triangular to a Lorentzian WL-peak was observed for increasing temperatures (there the 'truncation' of paths is imposed by the temperature-dependent mean free paths for inelastic or dephasing collisions, $L_{\rm max}=l_i$).\\ 
%
\section{Conclusions}
We have unambiguously identified pseudo-paths which are diffractively backscattered from the dot openings to be essential for the WL correction in the circular quantum dot. Our approach neither involves any semiclassical limit such as $\hbar\rightarrow 0$ (or $N\rightarrow\infty$) nor any assumptions about chaoticity or ensemble average. We believe that the diffractive contributions identified above provide an essential clue to the puzzle on WL at small lead mode numbers and in regular systems. The contributions result from short paths well below the Ehrenfest time $\tau_{E}$ \cite{BroRah06,JacWhi06,HeuMulBraHaa06} (unless diffractive corrections are included in the definition of $\tau_E$\cite{AigRotBur05}). Our results are in line with experiments \cite{ChaBarPfeWes94} in which WL is observed for moderate values of wavenumbers $k$ with only a few modes $N$ open and a phase-decohering mean free path $l_i$ extending only over a few times the diameter of the structure. The latter precludes significant contributions to the WL from $L>l_i$. We therefore suggest that diffraction plays a dominant role for finite $\hbar$ in chaotic systems as well. This raises interesting questions regarding the relation between our results and previously proposed theories for chaotic dots in the $\hbar\rightarrow 0$ limit.\cite{RicSie02,BroRah06,HeuMulBraHaa06,JacWhi06} In the present case, diffraction originates from the sharp edges at the lead. As long as the de Broglie wavelength $\lambda_D$ remains larger than the radius $r_c$ of the curvature of the rounded corners diffraction persists. (In Ref.~\onlinecite{ZozBer96} it was found that the amplitude of WL does not change upon rounding the lead edges for small mode numbers.) For $\lambda_{D}\ll r_c$, the quasi-classical electron 'sees' a convex-shaped boundary and the classically regular billiard turns chaotic. The limit $\hbar\rightarrow 0$ thus constitutes the crossover between strong contributions from diffraction to emergence of chaotic paths scattered at the rounded corners. We conclude that the present theory complements rather than contradicts previous models for the $\hbar\rightarrow0$ limit.\\ 
%
\section*{Acknowledgments}
We thank F.~Aigner, P.~Brouwer, F.~Libisch, S.~Rahav and A.~D.~Stone for helpful discussions. Work supported by the FWF Austria (Grant Nos.~SFB016 and P17359), the Max-Kade and the W.~M.~Keck foundations, and the French Partenariat Hubert Curien "Amadeus".


\begin{thebibliography}{99}

\bibitem{ErbLenMar93}
E. Akkermans, P.E. Wolf and R. Maynard, Phys. Rev. Lett. {\bf 56},  1471 (1986);
F.A. Erbacher, R. Lenke, and G. Maret, Europhys. Lett. {\bf 21}, 551 (1993).

\bibitem{deRTouDerRouFin05}
J. de Rosny, A. Tourin, A. Derode, P. Roux, and M. Fink, Phys. Rev. Lett. {\bf 95}, 074301 (2005)

\bibitem{CheSuLuHua06}
Y.F. Chen, K.W. Su, T.H. Lu, and K.F. Huang, Phys. Rev. Lett. {\bf 96}, 033905 (2006)

\bibitem{LarMar04}
E. Larose, L. Margerin, B. A. van Tiggelen, and M. Campillo, Phys. Rev. Lett. {\bf 93}, 048501 (2004).

\bibitem{KuhMinDelSigMul05}
R.C. Kuhn, C. Miniatura, D. Delande, O. Sigwarth, and C.A. M\"uller, Phys. Rev. Lett. {\bf 95}, 250403 (2005).

\bibitem{Ber84}
G. Bergmann, Phys. Rep. {\bf 107}, 1 (1984); S. Chakravarty and A. Schmid, {\it ibid.} {\bf 140}, 193 (1986).

\bibitem{GorLarKhm79}
L. P. Gorkov, A. I. Larkin, and D. E. Khmelnitskii, Pis'ma Zh. Eksp. Teor. Fiz. {\bf 30}, 248 (1979) [JETP Lett. {\bf 30}, 228 (1979)].

\bibitem{ChaBarPfeWes94}
A.M. Chang, H.U. Baranger, L.N. Pfeiffer, and K.W. West,
Phys. Rev. Lett. {\bf 73}, 2111 (1994).

\bibitem{BerKatMarWesGos94}
M. Berry, J. Katine, C. Marcus, R. Westervelt, and A.C. Gossard, Surf. Sci {\bf 305}, 495 (1994).

\bibitem{BarJalSto93}
H.U. Baranger, R.A. Jalabert, and A.D. Stone,
Phys. Rev. Lett. {\bf 70}, 3876 (1993); Chaos {\bf 3}, 665 (1993). 

\bibitem{BarMel94}
H.U. Baranger and P.A. Mello, Phys. Rev. Lett. {\bf 73}, 142 (1994).

\bibitem{AkiFerBirVas99}
R. Akis, D.K. Ferry, J.P. Bird, and D. Vasileska, Phys. Rev. B {\bf 60}, 2680 (1999).

\bibitem{AleLar96}
I.L. Aleiner and A.I. Larkin, Phys. Rev. B {\bf 54}, 14423 (1996).

\bibitem{RicSie02}
K. Richter and M. Sieber, Phys. Rev. Lett. {\bf 89}, 206801 (2002).

\bibitem{TakNak97}
Y. Takane and K. Nakamura, J. Phys. Soc. Jpn. {\bf 66}, 2977 (1997).

\bibitem{BroRah06}
P.W. Brouwer and S. Rahav, Phys. Rev. B {\bf 74}, 075322 (2006).


\bibitem{HeuMulBraHaa06}
S. Heusler, S. M\"uller, P. Braun, F. Haake, Phys. Rev. Lett. {\bf 96}, 066804 (2006).


\bibitem{JacWhi06}
Ph. Jacquod and Robert S. Whitney, Phys. Rev. B {\bf 73}, 195115 (2006).

\bibitem{BloZoz01} 
T. Blomquist and I. V. Zozoulenko, Phys. Rev. B {\bf 64}, 195301 (2001).

\bibitem{YanIshBur95}
X. Yang, H. Ishio, and J. Burgd\"orfer, Phys. Rev. B {\bf 52}, 8219 (1995).

\bibitem{RotTanWirTroBur00}
S. Rotter {\it et al.}, Phys. Rev. B {\bf 62}, 1950 (2000); {\bf 68}, 165302 (2003).

\bibitem{StaRotBurWir05}
L. Wirtz, C. Stampfer, S. Rotter, and J. Burgd\"orfer, Phys. Rev. E
{\bf 67}, 016206 (2003); C. Stampfer, S. Rotter, J. Burgd\"orfer, and L. Wirtz,
{\it ibid.} {\bf 72}, 036223 (2005).

\bibitem{SchAlfDel96}
C. D. Schwieters, J. A. Alford, and J. B. Delos, Phys. Rev. B
{\bf 54}, 10652 (1996).
 
\bibitem{WirTanBur97} 
L. Wirtz, J.-Z. Tang, and J. Burgd\"orfer, Phys. Rev. B
{\bf 56}, 7589 (1997).

\bibitem{SiePavSch97}
M. Sieber, N. Pavloff, and C. Schmit, Phys. Rev. E {\bf 55}, 2279 (1997).

\bibitem{KouPat74} 
R.~Kouyoumjian, P.~H.~Pathak, Proc.~IEEE {\bf 62}, 1448 (1974)

\bibitem{Kel62}
J.B. Keller, J. Opt. Soc. Amer. {\bf 52}, 116 (1962).

\bibitem{hik81}
S.~Hikami, Phys. Rev. B, {\bf 24}, 2671 (1981).

\bibitem{BreStaWirRotBur}
I.~B\v rezinov\'a et al.~(unpublished).

\bibitem{AigRotBur05}
F. Aigner, S. Rotter, and J. Burgd\"orfer, Phys. Rev. Lett. {\bf 94}, 216801 (2005);
S. Rotter, F. Aigner, and J. Burgd\"orfer, Phys. Rev. B {\bf 75}, 125312 (2007).

\bibitem{ZozBer96}
I.V. Zozoulenko and K.-F. Berggren, Phys. Rev. B {\bf 54}, 5823 (1996).

\end{thebibliography}
\end{document}